# Sub-ppb $CO_2$ Detection based on Dissipative Whispering Gallery mode Microcavity Sensor

Shujing Ruan[1#], Guangzhen Gao[1#], Jianing Zhang[1#], Haotian Wang[1*], Dongxing Cheng[1], Jun Guo[1], Chuanyong Ren[1], Weidong Chen[2], Deyuan Shen[1] & Tingdong Cai[1*]

[1]College of Physics and Electronic Engineering, Jiangsu Normal University, Xuzhou, 221116, China.
[2]Laboratoire de Physicochimie de l'Atmosphère, Université du Littoral Côte d'Opale 189A, 59140 Dunkerque, France.

#These authors contributed equally to this work.

*Corresponding author: e-mail: wanghaotian@jsnu.edu.cn (Haotian Wang), caitingdong@126.com (Tingdong Cai).

## Abstract

Whispering gallery mode (WGM) microcavities feature ultrahigh Q-factors and small mode volumes, offering strong light-matter interactions for sensing applications. However, unmodified surfaces are weakly responsive togas-phase refractive index changes, limiting trace gas detection. In this work, we propose a novel dissipative sensing scheme based on a non-functionalized WGM microcavity and experimentally demonstrate its feasibility and performance through trace-level $CO_2$ detection. Unlike conventional dispersive sensing that tracks resonance frequency shifts, our approach leverages enhanced local fields and thermal effects at the coupling region to convert weak absorption into measurable variations in resonance depth. A modulation-demodulation method suppresses low-frequency noise, with parameters optimized experimentally. The sensor achieves quantitative detection over 1.5-400 ppm ($R^2 > 0.99$), ~1.13% accuracy, a minimum detection limit of 0.12 ppb at 424 s integration-five orders of magnitude better than dispersive WGM sensors. Long-term $CO_2$ monitoring further demonstrates stability and resistance to environmental perturbations. Compared with state-of-the-art cavity-enhanced and photoacoustic sensors, our system delivers at least an order of magnitude lower baseline detection while maintaining compact size, ambient-pressure operation, and low cost, highlighting its potential for scalable, real-world deployment.

## Introduction

Precise and real-time measurement of gas concentration is of great significance in various fields, including environmental monitoring[1], industrial process control[2], public safety[3], and medical diagnostics[4]. With growing concerns over air pollution, greenhouse gas emissions, and toxic gas leaks, there is an increasing demand for gas sensing technologies that offer high sensitivity, high selectivity, miniaturization, and excellent stability[5].

Currently, gas detection technologies can be broadly classified into three categories: electrochemical and semiconductor material–based sensing methods[6-8], spectroscopic absorption–based methods[9,10], and emerging techniques based on photoacoustic effects or optical microstructures[11-13]. Electrochemical[14] and metal oxide semiconductor sensors[15] are widely adopted in household safety monitoring and smart devices due to their simple structure, fast response, and low manufacturing cost; however, their poor stability, susceptibility to temperature and humidity variations, limited selectivity, and significant cross-sensitivity constrain their use in complex or high-precision scenarios. Spectroscopic absorption techniques such as Tunable Laser Absorption Spectroscopy (TLAS)[16] and Non-Dispersive Infrared (NDIR)[17] exploit the specific absorption lines of gas molecules at selected wavelengths, providing high sensitivity and strong selectivity for atmospheric monitoring and industrial emissions measurement. Absorption spectroscopy aims to enhance light–molecule interaction. To this end, mid-infrared light sources are preferred, since molecular absorption cross-sections in this band exceed those in the near-infrared by one to three orders of magnitude, thereby significantly boosting sensitivity. High sensitivity also depends on a long effective optical path, which is typically achieved with multi-pass cells or integrated cavity. These configurations inevitably increase instrument size, structural complexity, and cost, all of which undermine portability. Moreover, laser-based absorption methods require stringent frequency stability to ensure quantitative accuracy and reproducibility. Locking the laser emission to a reference gas cell or a high-precision wavemeter—keeping frequency drift below 1 MHz—is essential to avoid measurement errors caused by laser frequency shifts.

In recent years, a number of emerging gas sensing techniques have been developed to overcome the limitations of traditional methods. These include photoacoustic[18,19] and photothermal[20,21] detection, fiber interferometry[22], fiber grating[23], and microcavity-based resonant sensing[24-26]. Most of them exploit interactions between gas molecules and light or sound, and offer key advantages such as compactness, low power consumption, and high sensitivity. Among these, optical microcavities based on whispering gallery modes (WGM) have shown great promise[27-30]. WGM cavities can be scaled down to the micrometer level while maintaining high Q-factors and excellent integrability. Thanks to their ultra-high-quality factors and small mode volumes, WGM sensors are exceptionally sensitive to slight changes in refractive index or absorption. Usually, WGM microcavities have dispersive sensing mechanism that detect the wavelength shift of microcavity resonance caused by changes in external parameters[31,32]. However, when the WGM microcavity sensor is applied to gas detection, the change in the refractive index of the microcavity caused by the change in the gas state is very weak.

Functional coatings, such as polymers, graphene, or metal-organic frameworks (MOFs), enhance selectivity by selectively interacting with specific gases like ammonia, carbon dioxide, and volatile organic compounds (VOCs)[33-38]. In gas sensing, functional coatings in WGM-based gas sensing face several limitations. They are susceptible to environmental factors such as temperature and humidity, which can cause interference and reduce detection accuracy. Over time, coatings may degrade due to chemical interactions or environmental exposure, requiring frequent maintenance. While designed for specificity, cross-sensitivity to similar analytes in complex mixtures can compromise selectivity. Additionally, the fabrication of effective coatings involves complex processes and high costs. The finite absorption capacity of these coatings limits their performance in high-concentration scenarios. These drawbacks highlight the need for more robust materials and advanced techniques to improve reliability and long-term usability in gas sensing applications. More importantly, the surface functionalization process of the WGM cavity will greatly deteriorate the Q factor, thereby reducing the sensitivity and detection limit of the sensor. Moreover, the dispersive WGM sensor requires the laser to have good frequency stability. The frequency fluctuation of the laser itself can cause the WGM sensor to misjudge whether the resonance wavelength shift is caused by the changes that need to be detected. And after long-term operation, the frequency drift of the laser will move away from the WGM cavity resonance region, causing the sensor function to fail. In order to stabilize the laser frequency, a stable frequency reference and a complex electronic feedback control system are usually used, which will inevitably increase the complexity and cost of the laser system[39,40].

Recently, WGM microcavities sensors using dissipative sensing mechanisms have attracted widespread attention. Dissipative WGM sensors monitor the changes in the transmitted light intensity at the microcavity resonance frequency due to the changes in the microcavity intrinsic or coupling loss caused by the external environment[41]. It does not need to measure the resonant wavelength shift as in the dispersive WGM sensor, thus avoiding the influence of laser frequency fluctuations on the sensor measurement results. This dissipative WGM microcavity sensor has been successfully applied to achieve high-sensitivity detection of sound waves, which relies on the mechanical vibration to cause changes in the coupling strength between the tapered fiber and the microcavity[42-44]. When the dissipative sensor is applied to gas detection, the absorption of light by the gas will cause the temperature near the microcavity to change. It will not only change the resonance wavelength of the microcavity, but also change the coupling strength of the light intensity inside the cavity to the outside. Therefore, it is only necessary to measure the change in transmittance at the resonance of the microcavity to achieve the inversion of gas concentration information, thereby overcoming the problem that the microcavity resonance frequency is insensitive to changes in the gas state. In particular, the light intensity enhanced by the high Q factor WGM microcavity will change the coupling strength more significantly due to gas absorption, so a more sensitive dissipative sensor can be obtained.

In this article, we propose a non-functionalized dissipative WGM gas sensor and demonstrate its performance using $CO_2$ gas concentration detection as an example. A silica microsphere WGM cavity is used as the sensing platform, and its resonance mode is excited by a tapered fiber. Notably, the field enhancement effect within the microcavity results in more intense heating in the coupling region. This ultimately causes a significant variation in the microcavity coupling strength, making the system highly sensitive to changes in $CO_2$ gas concentration. By establishing the relationship between the transmitted light intensity at the resonance and the gas concentration, the accurate measurement of $CO_2$ gas concentration can be achieved. The high Q-factor WGM microcavity not only amplifies the effect of heat generation due to gas absorption on the coupled system, but also the dissipative sensing mechanism eliminates the interference of laser frequency fluctuations on the measurement accuracy, and combines the modulation and demodulation technology to isolate the low-frequency intensity fluctuations of the laser. Our WGM microcavities have exceptionally high intracavity field enhancement, however, compensates for the inherently weak gas absorption in this spectral region, enabling gas detection with significantly improved sensitivity. The high-sensitivity sensor demonstrated in this work can achieve performance comparable to those using mid-infrared lasers even when working in the near-infrared band (1572.3nm), and finally the detection limit of $CO_2$ gas concentration can reach the ppt level.

## Results

**System configuration**

The WGM microcavity was fabricated by melting a tapered region of a single-mode silica optical fiber using localized electric discharge. First, we taper the fiber to about 10 μm in diameter, and then cut it off at the waist to form a tip. Through several electrode discharges, the fiber tip is melted to form a smooth microsphere with a diameter of about 60 μm, as illustrated in Fig. 1c.

The experimental setup for $CO_2$ sensing based on the dissipative WGM sensor is shown in Fig. 1a. The microsphere was affixed to a stable platform, while the tapered fiber was mounted onto a glass slide using UV-curable adhesive. The slide was placed on a three-axis micromanipulator to finely adjust the coupling between the tapered fiber and the microsphere, enabling optimal transmission of the WGM. A microscopic image of the coupled system is shown in Fig. 1c. The tapered fiber is attached to the microsphere surface to ensure coupling stability, and this coupling method can effectively enhance the sensitivity of the proposed dissipative sensor, which has been analyzed above. To allow controlled gas exchange, the fiber–microsphere coupling system was enclosed in a sealed chamber. Light coupling was achieved using a distributed feedback (DFB) tunable diode laser (NEL, NLK1L5GAAA) centered at ~1572.3 nm (Fig. 1e), corresponding to the $CO_2$ absorption line at 6359.967 cm$^{-1}$ (line strength: $1.761 \times 10^{-23}$ cm$^{-1}$/mol·cm$^2$). No significant interference from water vapor absorption is observed near this wavelength. The selected WGM resonance aligns closely with the absorption line center and has a much narrower linewidth, minimizing the impact of frequency jitter on signal strength. Lorentzian fitting of the resonance yields a linewidth of 0.00257 cm$^{-1}$, corresponding to a $Q$ factor of $2.47 \times 10^6$.

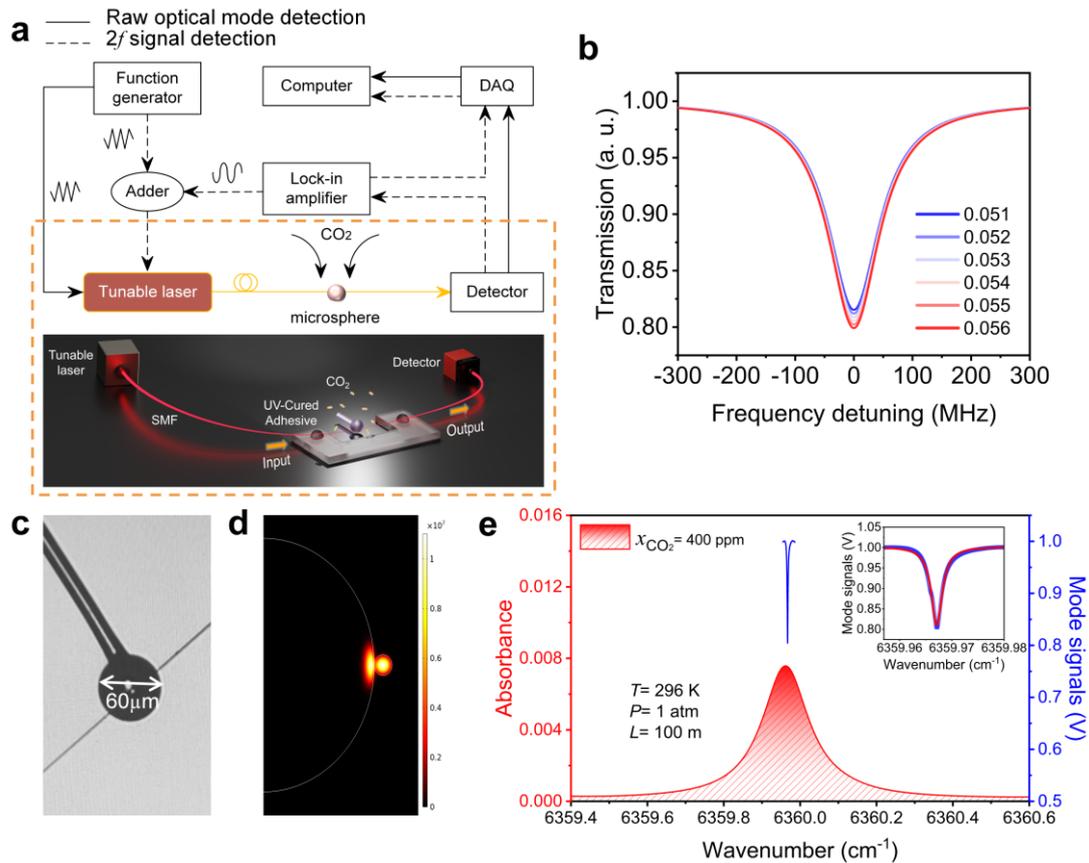

**Fig.1 Principle and implementation of dissipative sensing in a WGM microcavity.** **a**, Schematic diagram of the dissipative WGM gas sensor. **b**, Calculated transmitted light intensity near WGM resonance frequency at different $\kappa_{ext}/\kappa_{int}$. **c**, Microscopic photo of the coupling between 60 μm microsphere cavity and the tapered fiber. **d**, Mode field distribution of microsphere cavity and tapered fiber at coupling cross section. **e**, $CO_2$ absorption line and one resonance position of the microsphere cavity. Inserted graph: Lorentz profile fitting of the resonance mode.

**Optimization of modulation frequency and modulation amplitude**

The noise floor of the developed sensor mainly comes from the relative intensity noise (RIN) of the laser and the noise generated by the spontaneous temperature fluctuation of the WGM microcavity. The thermal noise of the microcavity is negligible compared to the RIN of the laser, which refers to a detailed analysis in Supplementary Note 3. Fig. 2a shows the measured noise floor in the 10Hz-100kHz frequency range, which include the RIN of the laser and the background noise of the detection and data acquisition system. It can be found that before 5kHz, the laser and detection system have excess noise. In order to eliminate the influence of this part of noise on the detection limit of the sensor, we will use modulation and demodulation technology to move the detection signal to 16 kHz to eliminate the low-frequency noise of the system.

During the signal modulation and demodulation process, the parameters of the sinusoidal waveform used for modulation—particularly the modulation frequency and depth—are critical to overall signal quality. To determine the optimal values of these parameters, we conducted $CO_2$ measurements in ambient air using the described system. The $2f$ signals corresponding to the selected optical mode were analyzed as a function of modulation frequency and modulation depth. Figs. 2b and 2c show the dependence of the $2f$ signal amplitude on modulation frequency and depth, respectively. Based on these results, a modulation frequency of 16.19 kHz and a modulation depth of 0.158 cm$^{-1}$ were selected for use in this study. As shown in the noise characterization data in Fig. 2a, the choice of 16.19 kHz effectively suppresses low-frequency noise from the laser and detection system, further validating its suitability.

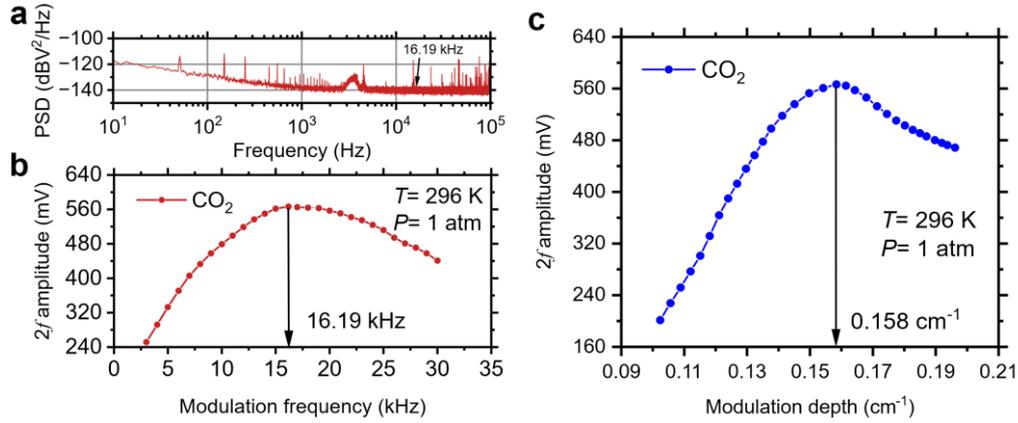

**Fig.2 Optimization of wavelength modulation spectroscopy parameters.** **a**, Noise power spectral density of laser and detection system in the 10 Hz–100 kHz frequency range. **b**, Variation of 2$f$ signal amplitude with modulation frequency. **c**, Variation of 2$f$ signal amplitude with modulation depth.

To assess the system calibration and measurement accuracy of the dissipative WGM gas sensor for $CO_2$ detection, we tested ten $CO_2$/$N_2$ mixtures with concentrations ranging from 1.5 to 400 ppm. The mixtures were dynamically generated by diluting a certified 1% $CO_2$ gas standard (in $N_2$) using a commercial gas mixing system. To assess the enhancement in detection performance enabled by modulation-demodulation techniques, both the raw optical mode signal (without modulation) and the demodulated second harmonic (2$f$) signal were acquired under each concentration condition. All measurements were performed at 1 atm and 296 K to ensure consistent experimental conditions. For each concentration, 20 independent signal sets were acquired within 7 minutes. Each set was the average of 200 repeated scans, improving the signal-to-noise ratio (SNR) and minimizing random errors to ensure reproducibility.

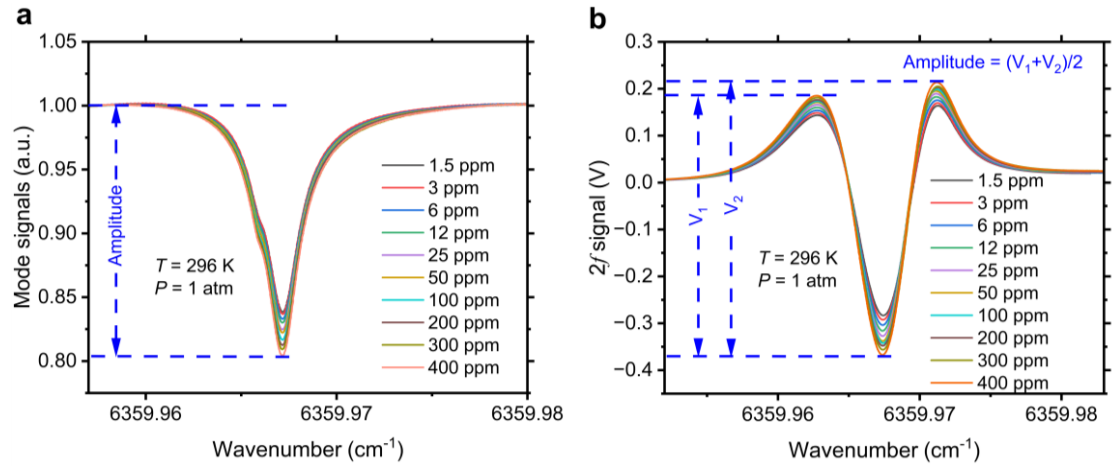

**Fig.3 Sensor response to varying CO$_2$ concentrations. a**, The raw optical mode signals measured at different CO$_2$ concentrations. **b**, The 2$f$ signals measured at different CO$_2$ concentrations.

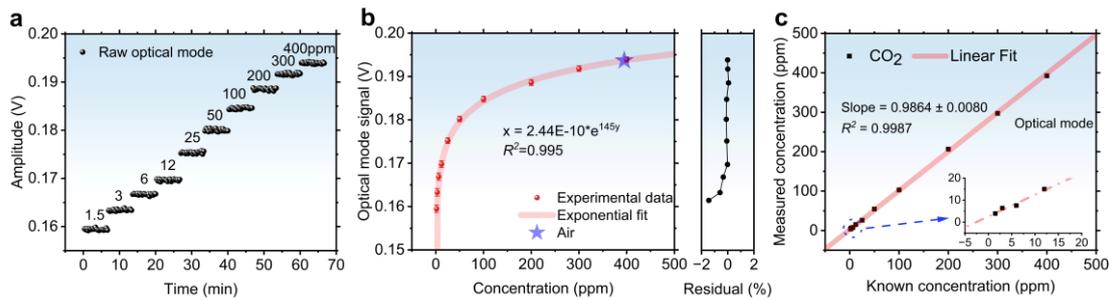

**Fig.4 Quantitative CO$_2$ sensing via dynamic response and calibration. a**, The measured optical mode amplitude versus calibration time for different CO$_2$ concentration levels ranging from 1.5 to 400 ppm. **b**, Experimental data points and the corresponding fitted curve of optical mode amplitude as a function of CO$_2$ concentration, showing residuals within 2%. **c**, Comparison between the CO$_2$ concentration measured by the raw optical mode and the nominal concentration recorded during gas preparation.

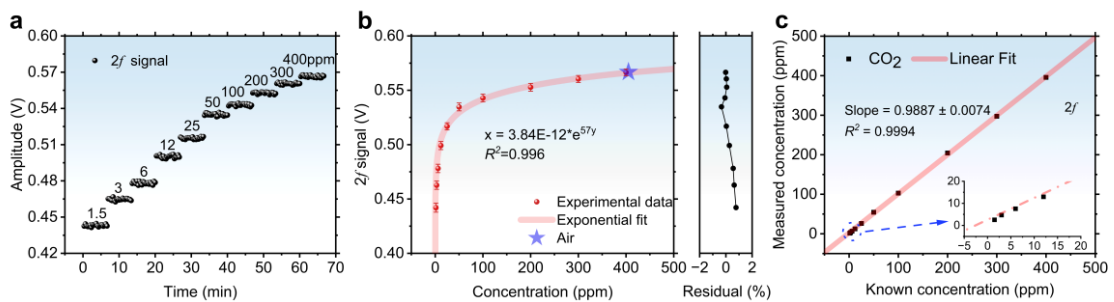

**Fig.5 High-precision calibration via 2$f$ signal. a**, The measured 2$f$ amplitude versus calibration time for different CO$_2$ concentration levels ranging from 1.5 to 400 ppm. **b**, Experimental data points and corresponding fitted curve of 2$f$ amplitude as a function of CO$_2$ concentration, showing residuals

within 1%. **c**, Comparison between the $CO_2$ concentration measured by 2*f* signal and the nominal concentration recorded during gas preparation.

**$CO_2$ concentrations measurement**

Figs. 3a and 3b present the raw signals for the optical mode of the microcavity and corresponding 2*f* signals at different $CO_2$ concentrations. It also can be seen from the figure that both signals exhibit high SNRs across the entire tested range (1.5–400 ppm), with peak amplitudes increasing monotonically with $CO_2$ concentration. According to equation (1) and the numerical simulation results in Fig. 1b, when the WGM microcavity is at under-coupled, the increase in coupling strength will lead to a decrease in the transmittance at the resonance. As the $CO_2$ gas concentration increases, the light absorption increases, which causes the tapered fiber to heat up and expand, thereby increasing its coupling strength to the WGM microcavity. The amplitude of the raw optical mode and 2*f* signals measured over a 70-minute period at different $CO_2$ concentrations is shown in Figs. 4a and 5a, respectively. To analyze the concentration-dependent behavior, the average signal amplitudes at each concentration level were plotted in Figs. 4b and 5b, where the vertical axis represents the amplitudes of the optical mode and 2*f* signal, and the horizontal axis corresponds to the target concentrations recorded during gas mixing. The fitting residuals, shown in the figures, are within 2% for the raw optical mode and 1% for the 2*f* signal.

According to the measured data, the signals follow an exponential relationship with $CO_2$ concentration. The fitting of the experimental data yielded the following equations, with correlation coefficients of $R^2 = 0.995$ and $R^2 = 0.996$ for the optical mode and 2*f* signal, respectively:

$$X_{CO_2} = 2.44\text{E-}10 * e^{145 P_{Mode}}, \quad X_{CO_2} = 3.84\text{E-}12 * e^{57 P_{Mode-2f}},$$

where $P_{Mode}$ and $P_{Mode-2f}$ represents the average of 100 peak values at each concentration level. The excellent fitting performance confirms the system's robust quantitative capability across a wide concentration range.

Figs. 4c and 5c compare the measured concentrations with reference values. Both signals exhibit strong linearity, with correlation coefficients of $R^2 = 0.9987$ (optical mode) and $R^2 = 0.9994$ (2$f$). The slopes of the linear fits are $0.9864 \pm 0.0080$ and $0.9887 \pm 0.0074$, corresponding to measurement accuracies of ~0.81% and ~0.75% for the optical mode signal and the 2$f$ signal, respectively. We further evaluated the system's practical applicability by measuring ambient $CO_2$ concentrations, which were determined to be 395 ppm and 405 ppm using the raw optical mode and the second-harmonic (2$f$) signal, respectively. The primary sources of error include uncertainties in gas mixing, imperfect peak extraction, laser frequency instability, and limitations in environmental pressure and temperature control accuracy.

**Measurement Precision and Detection Sensitivity**

To further evaluate the system's measurement precision, detection sensitivity, and long-term stability, continuous monitoring of a $CO_2$-$N_2$ mixture with a concentration of 1.5 ppm was conducted over a 1000 s period. The measured raw optical modes and 2$f$ signals are partially shown in Fig. S4 of the supplementary material. The time series of $CO_2$ concentrations retrieved from both the raw optical modes and 2$f$ signals and the corresponding frequency distribution histograms are shown in Figs. 6a and 6b. Both datasets are well fitted by Gaussian functions, indicating that the dominant noise source is white noise. The half-width at half-maximum (HWHM) of the Gaussian curves is 42.05 ppb and 10.79 ppb for the raw optical mode and 2$f$ signals, respectively, reflecting the practical measurement precision of each method. Notably, the modulation-demodulation technique effectively suppresses low-frequency noise, resulting in a ~4-fold improvement in measurement precision compared to the raw optical mode.

In addition, Allan deviation analysis was performed to assess the noise characteristics and theoretical detection limits under varying integration times. Fig. 6c show the Allan deviation curves derived from continuous 1.5 ppm $CO_2$ measurements using the raw optical mode and $2f$ signals, respectively. In both cases, noise levels decrease steadily with increasing integration time. At an integration time of 4 s, the minimum detection limits were determined to be 54.7 ppb (raw optical mode) and 17.11 ppb ($2f$ signal). These limits further decreased to 0.8 ppb and 0.12 ppb, respectively, when the integration time was extended to ~424 s, demonstrating the system's excellent detection sensitivity and long-term measurement stability. To further verify the reliability of the detection limit obtained from Allan variance analysis, measurements were performed on $CO_2$–$N_2$ mixtures with concentrations of 0.2 ppb and 0.5 ppb, prepared through multiple dilutions, under an integration time of 424 s. The corresponding raw optical mode signals and $2f$ signals are shown in Figs. 6d and 6e, respectively. The two ultra-low concentration levels can be clearly distinguished, demonstrating that the system maintains excellent detection sensitivity and stability even under near-limit conditions.

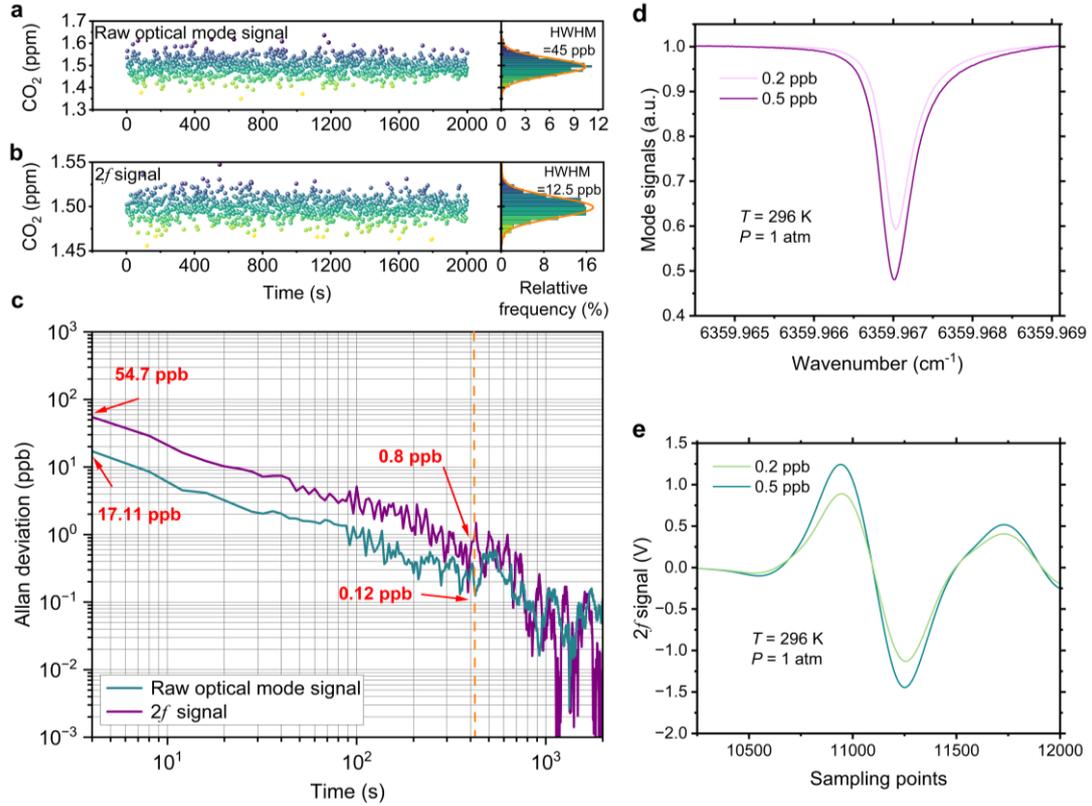

**Fig.6 Sensor stability and sub-ppb detection capability.** **a**, Distribution of $CO_2$ concentration measured continuously for 1000 s for 1.5 ppm $CO_2$-$N_2$ sample using raw optical mode ($T$=296 K, $P$=1 atm), as well as the frequency distribution and Gaussian profile fitting of measured concentration. **b**, Distribution of $CO_2$ concentration measured continuously for 16 minutes for 1.5 ppm $CO_2$-$N_2$ sample using $2f$ signal ($T$=296 K, $P$=1 atm), as well as the frequency distribution and Gaussian profile fitting of measured concentration. **c**, Allan deviation plot of the dissipative WGM gas sensor based on the data shown in Figs. 6a and 6b. **d** and **e,** Raw optical mode and $2f$ signals for $CO_2$-$N_2$ mixtures at 0.2 ppb and 0.5 ppb concentrations with 424 s integration time.

**Real-Time Monitoring of Ambient $CO_2$**

To evaluate the practical applicability of the system, long-term ambient $CO_2$ monitoring was conducted on the main campus of Jiangsu Normal University (JSNU, 34.19°N, 117.18°E, 10 m above ground level (a.g.l.)), as shown in Fig. 7a. The dissipative WGM-based $CO_2$ sensor was mounted on a compact mobile cart with physical dimensions of 110 × 65 cm² and a height of 0.9 m. To enable real-time monitoring of ambient air, the microsphere and the tapered fiber were removed from the sealed chamber and directly exposed to the environment. To minimize performance degradation caused by dust deposition, airborne particles, and strong air currents around the sensing region, a non-sealed black acrylic cover was placed above the microsphere and the tapered fiber, allowing airflow while providing basic protection.

The measured raw optical mode and 2*f* signals are partially shown in Fig. S5 of the supplementary material. Fig. 7b displays 48 h of $CO_2$ measurements from both the raw optical mode and 2*f* signals, revealing a clear diurnal cycle typical of a campus environment. From 00:00 to 06:00, $CO_2$ concentrations rise slowly as plant respiration continues, human activity remains minimal, and air movement is reduced. A rapid increase occurs between 06:00 and 08:00 with the start of morning activities. From 08:00 to 15:00, intensifying sunlight enhances photosynthesis and $CO_2$ uptake, driving concentrations downward to a daily minimum around midday. After 15:00, photosynthesis weakens as light fades, and $CO_2$ gradually climbs again due to resumed emissions. Finally, between 18:00 and 24:00, the cessation of photosynthesis combined with plant respiration, human presence, and traffic produces a pronounced nighttime peak. Importantly, the two independent 48-hour monitoring sessions conducted two days apart using both the raw optical mode and the 2*f* signal revealed highly consistent temporal trends, confirming the reliability and repeatability of the dissipative WGM-based $CO_2$ sensor. During sensor operation, the laser maintains frequency scanning mode. A strong field is established within the microsphere only when it approaches the WGM resonance. At this point, $CO_2$ absorption increases instantaneously, resulting in heat generation. Because the thermal expansion of the tapered fiber in the coupling region is instantaneous, it is not affected by slow thermal fluctuations in the external environment. Fig. S6 in the supplementary materials presents the raw mode signals measured for the same 400 ppm $CO_2$–$N_2$ sample by varying the temperature of the sealed chamber containing the WGM microcavity. As shown, the signals remain highly consistent across different temperatures, with deviations within 1% (detailed analysis is provided in Supplementary Note 5).

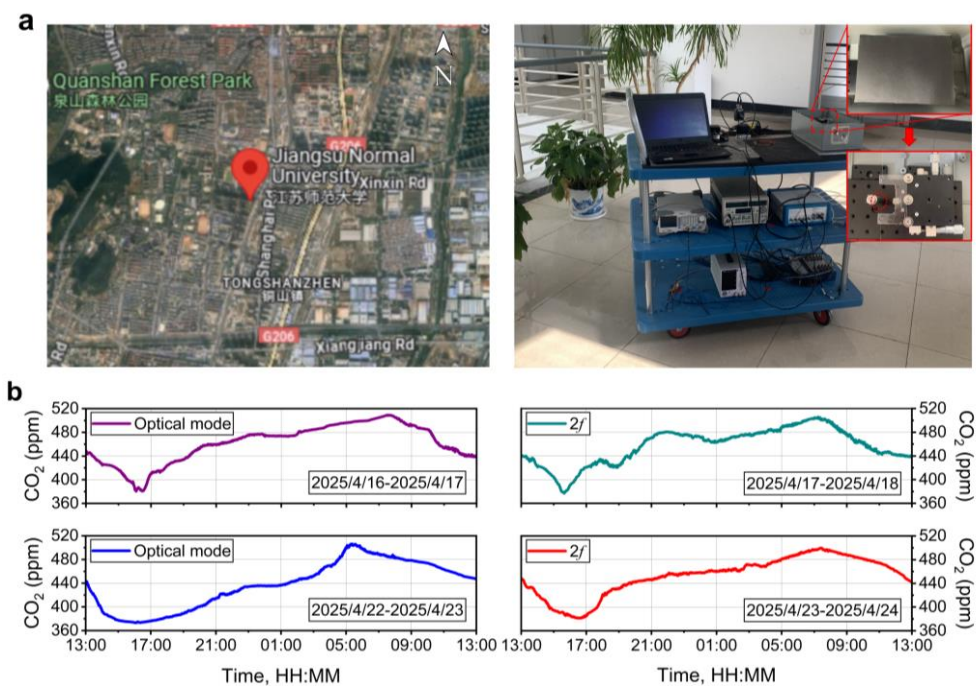

**Fig.7 Continuous atmospheric $CO_2$ monitoring in a real-world environment.** **a**, Detailed locations of dissipative WGM-based $CO_2$ sensor measurement site and its surroundings in Xuzhou. Map provided by Google Earth. **b**, Time series of $CO_2$ measured from outdoor air using the raw optical signal and $2f$ signal of the dissipative WGM gas sensor during~48h.

**Comparison with other methods**

The measurement performance of the sensor is systematically evaluated by comparison with several representative optical systems of similar functionality whose excitation sources span the near-infrared to mid-infrared regions. The comparison focuses on key parameters including sample cell volume, system cost, difficulty level of optical path adjustment, measurement precision, and limit of detection, as summarized in Fig. 8 and Table 1. The results indicate that the developed sensor achieves a detection limit as low as 0.12 ppb, representing an order of magnitude improvement over current mainstream mature systems. In terms of sample cell volume, optical alignment complexity, and overall system cost, the developed sensor exhibits significant advantages.

Specifically, the sample cell volume is only 20% of the smallest conventional design, and the construction cost of the cell is merely USD 0.1. Moreover, the system enables highly sensitive measurements under ambient pressure, thereby eliminating the need for low-pressure operation typically required by conventional spectroscopic techniques. This feature simplifies system sealing and reduces the associated engineering complexity and potential measurement inaccuracies. For a detailed comparison, refer to Table 1 in the supplementary material.

Compared with typical surface functionalized dispersive-based WGM microcavity systems, the present system demonstrates a roughly five orders of magnitude improvement in detection limit. Furthermore, it utilizes the modal peak height as the retrieval parameter for concentration, effectively mitigating the influence of mode drift induced by environmental temperature or pressure fluctuations, which is a common issue in dispersive systems. This approach significantly enhances long-term monitoring stability. Experimental results show that under temperature variations of $\pm 10$ K and pressure fluctuations of $\pm 50$ Pa, the concentration measurement deviation remains within $\pm 1\%$, whereas the deviation for dispersive systems can reach up to $\pm 5\%$.

The developed sensor not only maintains high measurement performance, but also features a compact structure, low cost, and strong anti-interference capability. Additionally, it benefits from the communication-band light sources and detectors, which offer high commercial maturity, stable performance, low cost, and ease of integration. These advantages contribute to the system's strong potential for engineering implementation and broad applicability in field measurements.

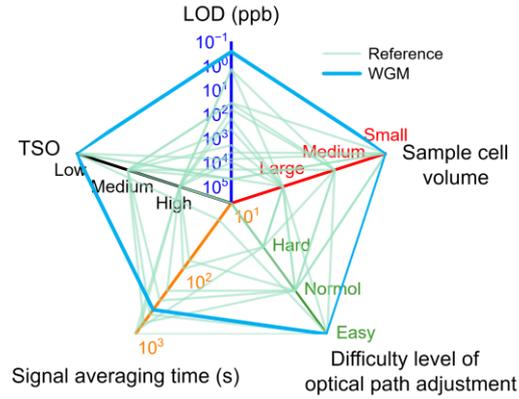

**Fig.8 Comparison of Sensor Performance.** Performance comparison of the dissipative WGM gas sensor with 20 state-of-the-art $CO_2$ sensors from the literature (Table 1 with Supporting References).

A comparison between the present system and other functionally similar systems in terms of key parameters, including difficulty level of optical path adjustment, sample cell volume, system cost, measurement accuracy, and detection limit.

**Table 1.** Comparison of gas sensing performance of 20 state-of-the-art $CO_2$ sensors

| Spectral technology | LOD[a] | Sample cell volume | Difficulty level of optical path adjustment | TSO[b] | Optimal integration time | Wavelength |
|---|---|---|---|---|---|---|
| Multi-pass cell[45] | 450 ppb | Large | Normol | Medium | 0.6 s | 2.004 μm |
| OA-CEAS[46] | 1.40 ppb | Large | Hard | High | 100 s | 1.6 μm |
| CRDS[47] | 160 ppb | Large | Hard | High | 816 s | 1.6 μm |
| On-chip[48] | 7.56 ppm | Small | Easy | Low | 18 s | 4.23 μm |
| SEIRA[49] | 20.0 ppm | Medium | Hard | Medium | N/A | 0.125 μm |
| PAS[50] | 37.0 ppb | Medium | Normol | Medium | 307 s | 4.3 μm |
| QEPAS[51] | 35.0 ppb | Small | Normol | Medium | 670 s | 2.0 μm |
| MEMS[52] | 50.0 ppm | Small | Easy | Low | N/A | 4.3 μm |
| CEPAS[53] | 4.00 ppm | Small | Normol | High | 75 s | 1.57 μm |
| Photothermal Spectroscopy[54] | 1.50 ppb | Medium | Easy | High | 1000 s | 2.003 μm |
| LITES[55] | 47.7 ppm | Large | Easy | Medium | 500 s | 1.57 μm |
| Dispersion Spectroscopy[56] | 65.78 ppb | Medium | Normol | High | 218 s | 4.329 μm |
| NDIR[57] | 2.00 ppm | Large | Normol | Medium | N/A | 4.3 μm |
| FTIR[58] | 10.0 ppm | Large | Hard | High | 5 s | 4.25 μm |
| Fiber FP Interferometer[59] | 480 ppb | Small | Easy | Medium | N/A | 4.266 μm |
| Fiber grating[60] | 401 ppm | Medium | Easy | Medium | 20 s | 700 nm |
| Silicon microring resonator[61] | 370 ppm | Small | Easy | High | N/A | 1560 nm |
| Antiresonant Fiber[62] | 144 ppm | Small | Normal | Medium | 1.5 s | 1574 nm |
| Photonic Crystals[63] | 2000 ppm | Small | Easy | Medium | N/A | 463-668 nm |
| Functionalized dispersed WGM[64] | 50.0 ppm | Small | Easy | Low | N/A | 1.525-1.57 μm |
| ***Dissipative WGM (This work)*** | ***0.12 ppb*** | ***Small*** | ***Easy*** | ***Low*** | ***424 s*** | ***1. 572 μm*** |

[a]LOD represents the limit of detection.
[b]TSO represents total system cost.

## Discussion

In this work, we developed a gas sensor based on a non-functionalized dissipative WGM microcavity, and demonstrated its performance using $CO_2$ detection in the 1.5 μm band as a representative example. A silica microsphere was employed as the sensing cavity. By selecting an optical mode aligned with the $CO_2$ absorption line, we analyzed the system performance using both the raw transmission signal and the 2$f$ signal obtained through modulation-demodulation techniques. Allan variance analysis revealed that, with a 4 s integration time, the minimum detection limits (MDLs) were 54.7 ppb and 17.11 ppb for the optical mode and 2$f$ signals, respectively. Extending the integration time to approximately 424 s further improved the minimum detection limits to 0.8 ppb and 0.12 ppb. The 2$f$ signal exhibited significantly better noise suppression and sensitivity, primarily due to its ability to reject low-frequency noise arising from laser intensity fluctuations and environmental disturbances. Comparison with state-of-the-art optical gas sensors confirms the excellent sensitivity and compactness of the proposed system. Benefiting from the high-Q factor and microscale volume of the WGM cavity, this sensor achieves high sensitivity while maintaining low cost, compact structure, and strong environmental robustness. In addition, the system supports ambient-pressure operation, eliminating the need for vacuum components and simplifying engineering deployment.

Due to the high Q factor of silica microspheres in the 1–2 μm spectral range and the use of a non-functionalized configuration, the proposed dissipative WGM-based sensing system is broadly adaptable for detecting various trace gases with near-infrared absorption features—including but not limited to [$CO_2$, $CH_4$, $NH_3$, $H_2S$, $C_2H_2$, and HF]—simply by selecting appropriate laser wavelengths. This versatility, combined with the advantages of telecom-grade light sources and detectors, positions the developed dissipative WGM-based sensor as a promising solution for high-sensitivity, compact, and low-cost trace gas detection in both laboratory and field environments.

## Methods

**Theory and simulations**

When a tapered fiber is used to couple laser into a WGM microcavity, the intensity transmittance of the tapered fiber at resonance can be described as:

$$\Gamma_0 = \left(\frac{\kappa_{ext} - \kappa_{int}}{\kappa_{ext} + \kappa_{int}}\right)^2 \quad (1)$$

where $\kappa_{int}$ and $\kappa_{ext}$ represent the loss rates due to the intrinsic loss in the WGM microcavity and the loss rate introduced by coupling (coupling strength), respectively (detailed analysis is provided in Supplementary Note 1). For our dissipative WGM sensor, the change in $\Gamma_0$ can be used as the sensor readout. For gas absorption sensors, the change in $\Gamma_0$ is mainly due to the contribution of $\kappa_{ext}$ rather than $\kappa_{int}$. This is because the intrinsic loss rate $\kappa_{int}$ caused by gas absorption is negligible. However, gas absorption also heats up the tapered fiber and causes it to expand. Especially, since the field intensity inside the high-Q WGM microcavity is much greater than the incident light, the tapered fiber in the coupling region (attached to the microcavity surface in the following experiment) can more sensitively sense the heating effect. It can not only affect the refractive index change of the tapered fiber and the microcavity, but also change the coupling strength $\kappa_{ext}$ caused by the thermal expansion of tapered fiber in coupling region. When $\kappa_{ext}$ changes, the change in $\Gamma_0$ can be

expressed as:

$$\frac{\delta \Gamma_0}{\Gamma_0} = \frac{1}{\Gamma_0}\frac{\partial \Gamma_0}{\partial \kappa_{\text{ext}}}\delta \kappa_{\text{ext}} \tag{2}$$

We can relate $\delta\kappa_{ext}$ to temperature change $\delta T$ through thermal expansion and the thermo-optic coefficient:

$$\frac{\delta \Gamma_0}{\Gamma_0} = \frac{4\kappa_{int}\kappa_{\text{ext}}}{|\kappa_{ext}^2 - \kappa_{int}^2|}\left(\frac{\partial \kappa_{\text{ext}}}{\partial l}l\alpha_l + \frac{\partial \kappa_{\text{ext}}}{\partial n}n\alpha_n\right)\delta T \tag{3}$$

where $l$ and $n$ represent the tapered fiber length and refractive index, respectively, $\alpha_l = \frac{1}{l}\frac{\partial l}{\partial T}$ and $\alpha_n = \frac{1}{n}\frac{\partial n}{\partial T}$ are the thermal expansion and thermo-optic coefficients of silica, respectively. The optical field distribution of the WGM microcavity and tapered fiber at coupling cross section is shown in Fig. 1d. The coupling strength depends on the mode field overlap ratio between tapered fiber and microcavity as well as the length of coupling region, and the latter is more sensitive to temperature changes (detailed analysis is provided in Supplementary Note 2).

We calculated the transmission spectrum of one WGM, as shown in Fig. 1b. In the calculation, we assumed that $\kappa_{int} = 2\pi \times 100$ MHz, which corresponds to the $Q$ factor of $\sim 2 \times 10^6$ for this WGM. When $\kappa_{ext}/\kappa_{int}$ changes from 0.051 to 0.056, so the range of $\kappa_{ext}$ is from 32 to 35 MHz, we found $\Gamma_0$ changes from 0.8 to 0.816. At this time, the WGM microcavity is at the under-coupled state ($\kappa_{ext} < \kappa_{int}$), and the increase in $\kappa_{ext}$ will reduce $\Gamma_0$. This is consistent with the experimental results (Fig. 3a): when the increase in gas concentration causes enhanced light absorption, we will observe a decrease in transmittance at resonance. It is worth noting that if the WGM microcavity is at the over-coupled state ($\kappa_{ext} > \kappa_{int}$), the increase in $\kappa_{ext}$ will increase $\Gamma_0$, but the sensitivity to the change in $\Gamma_0$ will also be reduced greatly. Therefore, in the experiment, the coupling position of the tapered fiber is adjusted to make the microcavity at the under-coupled state, thereby improving the sensitivity for our dissipative WGM sensor.

**Experimental details**

A laser driver (Stanford Research Systems, LDC501) controlled the temperature and current of the butterfly diode laser, which delivered an output power of ~10 mW. To perform $CO_2$ concentration measurements, both the raw optical mode signal and the second harmonic ($2f$) signal obtained via Modulation-Demodulation were acquired. When recording the raw signal, the butterfly diode laser's frequency was scanned by applying a 100 Hz, 300 mV triangular waveform from a function generator (Rigol DG1000Z). For $2f$ signal detection, this triangular drive was summed with the sinusoidal modulation output of a digital lock-in amplifier (Zurich Instruments MFLI) via an adder, and the combined waveform was applied to the laser. The output of the laser was coupled into the tapered fiber via a flange connector, and the transmitted light was detected by a photodetector (Thorlabs, PDA20CS2). For raw signal acquisition, the photodetector output was directly recorded by a data acquisition (DAQ) system (National Instruments, USB-6356). For $2f$ detection, the modulated signal was demodulated by the lock-in amplifier before DAQ. All signal acquisition, fitting, and peak extraction were automated via a LabVIEW-based program. $CO_2$ gas mixtures of varying concentrations were generated using a mass flow-controlled gas mixer (Environics, Series 4000), ensuring precise mixing and minimizing human error. During all measurements, the chamber environment was held constant at $T = 296$ K and $P = 1$ atm to avoid perturbations from temperature or pressure fluctuations.

# References

1. Zhao, X. *et al.* Titanium nitride sensor for selective $NO_2$ detection. *Nat. Commun.* **16**, 182 (2025).

2. Ydir, B. *et al.* Aluminum doped zinc oxide nanoplatelets based sensor with enhanced hydrogen sulfide detection. *Sci. Rep.* **15**, 8633 (2025).

3. Mandal, S. *et al.* A robust organic hydrogen sensor for distributed monitoring applications. *Nat. Electron.* **8**, 343-352 (2025).

4. Yang, L. *et al.* Moisture-resistant, stretchable NOx gas sensors based on laser-induced graphene for environmental monitoring and breath analysis. *Microsyst. Nanoeng.* **8**, 78 (2022).

5. Yuan, H., Li, N., Fan, W., Cai, H. & Zhao, D. Metal-organic framework based gas sensors. *Adv. Sci.* **9**, 2104374 (2022).

6. Gorbova, E. *et al.* Fundamentals and principles of solid-state electrochemical sensors for high temperature gas detection. *Catalysts* **12** (2021).

7. Williams, D. E. Semiconducting oxides as gas-sensitive resistors. *Sens. Actuators B Chem.* **57**, 1-16 (1999).

8. Zhao, X. *et al.* Titanium nitride sensor for selective $NO_2$ detection. *Nat. Commun.* **16**, 182 (2025).

9. Kosterev, A. A., Bakhirkin, Y. A., Curl, R. F. & Tittel, F. K. Quartz-enhanced photoacoustic spectroscopy. *Opt. Lett.* **27**, 1902-1904 (2002).

10. Li, C., Dong, L., Zheng, C. & Tittel, F. K. Compact TDLAS based optical sensor for ppb-level ethane detection by use of a 3.34μm room-temperature CW interband cascade laser. *Sens. Actuators B Chem.* **232**, 188-194 (2016).

11. Rosencwaig, A. & Gersho, A. Theory of the photoacoustic effect with solids. *J. Appl. Phys.* **47**, 64-69 (1976).

12. Wu, H. *et al.* Beat frequency quartz-enhanced photoacoustic spectroscopy for fast and calibration-free continuous trace-gas monitoring. *Nat. Commun.* **8**, 15331 (2017).

13. Palzer, S. Photoacoustic-based gas sensing: a review. *Sensors* **20** (2020).

14. Wan, H., Yin, H., Lin, L., Zeng, X. & Mason, A. J. Miniaturized planar room temperature ionic liquid electrochemical gas sensor for rapid multiple gas pollutants monitoring. *Sens. Actuators B Chem.* **255**, 638-646 (2018).

15. Ye, W. *et al.* Dimensional reduction in $Cs_2AgBiBr_6$ enables long-term stable Perovskite-based gas sensing. *Nat. Commun.* **16**, 4820 (2025).

16. Li, C., Dong, L., Zheng, C. & Tittel, F. K. Compact TDLAS based optical sensor for ppb-level ethane detection by use of a 3.34μm room-temperature CW interband cascade laser. *Sens. Actuators B Chem.* **232**, 188-194 (2016).

17. Tan, X. *et al.* Non-dispersive infrared multi-gas sensing via nanoantenna integrated narrowband detectors. *Nat. Commun.* **11**, 5245 (2020).

18. Qiao, S. *et al.* Ultra-highly sensitive dual gases detection based on photoacoustic spectroscopy by exploiting a long-wave, high-power, wide-tunable, single-longitudinal-mode solid-state laser. *Light Sci. Appl.* **13**, 100 (2024).

19. Wang, J. *et al.* Quartz-enhanced multiheterodyne resonant photoacoustic spectroscopy. *Light Sci. Appl.* **13**, 77 (2024).

20. Lin, Y. *et al.* Pulsed photothermal interferometry for spectroscopic gas detection with hollow-core optical fibre. *Sci. Rep.* **6**, 39410 (2016).

21. Zhao, P. *et al.* Mode-phase-difference photothermal spectroscopy for gas detection with an anti-resonant hollow-core optical fiber. *Nat. Commun.* **11**, 847 (2020).

22. Gong, Z., Chen, K., Yang, Y., Zhou, X. & Yu, Q. Photoacoustic spectroscopy based multi-gas detection using high-sensitivity fiber-optic low-frequency acoustic sensor. *Sens. Actuators B Chem.* **260**, 357–363 (2018).

23. Hromadka, J., Tokay, B., Correia, R., Morgan, S. P. & Korposh, S. Carbon dioxide measurements using long period grating optical fibre sensor coated with metal organic framework HKUST-1. *Sens. Actuators B Chem.* **255**, 2483–2494 (2018).

24. Mi, G., Horvath, C. & Van, V. Silicon photonic dual-gas sensor for $H_2$ and $CO_2$ detection. *Opt. Express* **25**, 16250-16259 (2017).

25. Li, H., Sun, B., Yuan, Y. & Yang, J. Guanidine derivative polymer coated microbubble resonator for high sensitivity detection of $CO_2$ gas concentration. *Opt. Express* **27**, 1991 (2019).

26. Gregor, M. *et al.* An alignment-free fiber-coupled microsphere resonator for gas sensing applications. *Appl. Phys. Lett.* **96**, 231102 (2010).

27. Needham, L.-M. *et al.* Label-free detection and profiling of individual solution-phase molecules. *Nature* **629**, 1062–1068 (2024).

28. Cai, L., Pan, J., Zhao, Y., Wang, J. & Xiao, S. Whispering gallery mode optical microresonators: structures and sensing applications. *Physica Status Solidi (a)* **217**, 1900825 (2020).

29. Toropov, N. *et al.* Review of biosensing with whispering-gallery mode lasers. *Light Sci. Appl.* **10**, 42 (2021).

30. Armani, D. K., Kippenberg, T. J., Spillane, S. M. & Vahala, K. J. Ultra-high-Q toroid microcavity on a chip. *Nature* **421**, 925-928 (2003).

31. Jiang, X., Qavi, A. J., Huang, S. H. & Yang, L. Whispering-gallery sensors. *Matter* **3**, 371–392 (2020).

32. Wang, Z., Zhou, B. & Zhang, A. P. High-Q WGM microcavity-based optofluidic sensor technologies for biological analysis. *Biomicrofluidics* **18**, 041502 (2024).

33. Passaro, V. M. N., Dell'Olio, F. & De Leonardis, F. Ammonia optical sensing by microring resonators. *Sensors* **7**, 2741–2749 (2007).

34. Sun, Y. *et al.* Optofluidic ring resonator sensors for rapid DNT vapor detection. *Analyst* **134**, 1386 (2009).

35. Sun, Y., Shopova, S. I., Frye-Mason, G. & Fan, X. Rapid chemical-vapor sensing using optofluidic ring resonators. *Opt. Lett.* **33**, 788 (2008).

36. Yao, B. *et al.* Graphene-enhanced brillouin optomechanical microresonator for ultrasensitive gas detection. *Nano Lett.* **17**, 4996–5002 (2017).

37. Li, C. *et al.* Part-per-Trillion trace selective gas detection using frequency locked whispering-gallery mode microtoroids. *ACS Appl. Mater. Interfaces* **14**, 42430–42440 (2022).

38. Cao, X., Yang, H., Wu, ZL. *et al.* Ultrasound sensing with optical microcavities. *Light Sci. Appl.* **13**, 159 (2024).

39. Kessler, T. *et al.* A sub-40-mHz-linewidth laser based on a silicon single-crystal optical cavity. *Nat. Photonics* **6**, 687-692 (2012).

40. Black, E. D. An introduction to Pound–Drever–Hall laser frequency stabilization. *American Journal of Physics* **69**, 79–87 (2001).

41. Jia, T. *et al.* Ultrahigh resolution acceleration sensing based on prism-microcavity dissipative coupling architecture. *ACS Photonics* **11**, 428–436 (2024).

42. Yu, D., Guo, X., Jiang, B. & Che, K. Acoustic modulation on a fiber laser based on dissipative coupling of a microcavity. *Opt. Express* **32**, 39065 (2024).

43. Meng, J.-W. *et al.* Dissipative acousto-optic interactions in optical microcavities. *Phys. Rev. Lett.* **129**, 073901 (2022).

44. Basiri-Esfahani, S., Armin, A., Forstner, S. & Bowen, W. P. Precision ultrasound sensing on a chip. *Nat. Commun.* **10**, 132 (2019).

45. Gu, M. *et al.* Open-path anti-pollution multi-pass cell-based TDLAS sensor for the online measurement of atmospheric $H_2O$ and $CO_2$ fluxes. *Opt. Express* **30**, 43961-43972 (2022).

46. Gao, G., Yang, Y., Wang, X., Yang, H. & Cai, T. FDM-assisted opposite two-way OA-CEAS system employing four lasers for simultaneous multi-species detection. *Opt. Express* **31**, (2023).

47. Xu, L. *et al.* Synchronous detection of $CO_2$ and $CH_4$ gas absorption spectroscopy based on TDM and optimized adaptive Whittaker-Henderson filtering algorithm. *Appl. Phys. B* **130** (2024).

48. Kim, I. *et al.* Freestanding germanium photonic crystal waveguide for a highly sensitive and compact mid-infrared on-chip gas sensor. *ACS Sensors* **9**, 5116-5126 (2024).

49. Chang, Y. *et al.* All-dielectric surface-enhanced infrared absorption-based gas sensor using guided resonance. *ACS Appl. Mater. Interfaces* **10**, 38272-38279 (2018).

50. Srivastava, A., Tian, Y., Bittner, A. & Dehé, A. Design and characterization of macroscopic indirect photoacoustic gas sensor. in *2022 IEEE Sensors*, pp.1-4 (2022).

51. Li, B., Wu, H., Feng, C., Jia, S. & Dong, L. Noninvasive skin respiration ($CO_2$) measurement based on quartz-enhanced photoacoustic spectroscopy. *Anal. Chem.* **95**, 6138-6144 (2023).

52. Vincent, T. A. & Gardner, J. W. A low cost MEMS based NDIR system for the monitoring of carbon dioxide in breath analysis at ppm levels. *Sens. Actuators B Chem.* **236**, 954-964 (2016).

53. Laurila, T. *et al.* Cantilever-based photoacoustic detection of carbon dioxide using a fiber-amplified diode laser. *Appl. Phys. B* **83**, 285-288 (2006).

54. Chen, F. *et al.* Frequency-division-multiplexed multicomponent gas sensing with photothermal spectroscopy and a single NIR/MIR fiber-optic gas cell. *Anal. Chem.* **94**, 13473-13480 (2022).

55. Hou, J. *et al.* Highly sensitive $CO_2$-LITES sensor based on a self-designed low-frequency quartz tuning fork and fiber-coupled MPC. *Chin. Opt. Lett.* **22**, 073001 (2024).

56. Zhang, H. *et al.* Measurement of $CO_2$ isotopologue ratios using a hollow waveguide-based mid-Infrared dispersion spectrometer. *Anal. Chem.* **95**, 18479-18486 (2023).

57. Tan, X. *et al.* Non-dispersive infrared multi-gas sensing via nanoantenna integrated narrowband detectors. *Nat. Commun.* **11**, 5245 (2020).

58. Scharf, T. *et al.* Gas detection with a micro FTIR spectrometer in the MIR region. *Procedia Chem.* **1**, 1379-1382 (2009).

59. Gong, Z., Chen, K., Yang, Y., Zhou, X. & Yu, Q. Photoacoustic spectroscopy based multi-gas detection using high-sensitivity fiber-optic low-frequency acoustic sensor. *Sens. Actuators B Chem.* **260**, 357-363 (2018).

60. Hromadka, J., Tokay, B., Correia, R., Morgan, S. P. & Korposh, S. Carbon dioxide measurements using long period grating optical fibre sensor coated with metal organic framework HKUST-1. *Sens. Actuators B Chem.* **255**, 2483-2494 (2018).

61. Mi, G., Horvath, C. & Van, V. Silicon photonic dual-gas sensor for $H_2$ and $CO_2$ detection. *Opt. Express* **25**, 16250-16259 (2017).

62. Jaworski, P. *et al.* Antiresonant hollow-core fiber-based dual gas sensor for detection of methane and carbon dioxide in the near- and mid-infrared regions. *Sensors* **20**, 3813 (2020).

63. Hong, W. et al. Full-color $CO_2$ gas sensing by an inverse opal photonic hydrogel. *Chem. Commun.* **49**, 8229 (2013).

64. Li, H., Sun, B., Yuan, Y. & Yang, J. Guanidine derivative polymer coated microbubble resonator for high sensitivity detection of $CO_2$ gas concentration. *Opt. Express* **27**, 1991-2000 (2019).

## Acknowledgements

We acknowledge the financial support from the National Natural Science Foundation of China (62275110, 42275136, 62305136), Funded by Basic Research Program of Jiangsu (BK20251926).

## Author contributions

H.T.W. and T.D.C. conceived the project. S.J.R., G.Z.G., J.N.Z., and D.X.C. carried out the experiments and data analysis. J.G. and C.Y.R. performed theory calculations and conducted simulations. W.D.C. and D.Y.S. provided valuable suggestions on data analysis and manuscript preparation. H.T.W. and T.D.C. were responsible for funding acquisition. All the authors contributed and discussed the content of this manuscript.

## Competing interests

The authors declare no competing interests.

## Additional information

**Supplementary information** The online version contains supplementary material available at

**Correspondence** and requests for materials should be addressed to Haotian Wang or Tingdong Cai.